\documentclass[iop]{emulateapj}
\usepackage{amsmath,graphics,epsfig}
\usepackage{natbib,hyperref}
\usepackage{soul}
\usepackage[usenames,dvipsnames]{xcolor} 
\usepackage{enumerate}

\newcommand{\bfH}{\ensuremath {\mathbf{H}}}
\newcommand{\bfk}{\ensuremath {\mathbf{k}}}

\newcommand{\bfell}{\ensuremath {\boldsymbol{\ell}}}

\newcommand{\bftheta}{\ensuremath {\boldsymbol{\theta}}}

\newcommand{\bfx}{\ensuremath {\mathbf{x}}}
\newcommand{\bfv}{\ensuremath {\mathbf{v}}}
\newcommand{\hatk}{\ensuremath {\mathbf{\hat k}}}
\newcommand{\hatH}{\ensuremath {\mathbf{\hat H}}}

\newcommand{\hattheta}{\ensuremath {\boldsymbol{\hat \theta}}}

\newcommand{\hata}{\ensuremath {\mathbf{\hat a}}}

\newcommand{\hatxi}{\ensuremath {\boldsymbol{\hat \xi}}}

\def\VEV#1{\left\langle #1 \right\rangle}

\begin{document}

\title{Dust-polarization maps and interstellar turbulence}

\shorttitle{Dust polarization and ISM turbulence}

\author{Robert R.\ Caldwell$^1$, Chris Hirata$^2$, and Marc
     Kamionkowski$^3$}

\affil{$^1$Department of Physics and Astronomy, 6127 Wilder
     Laboratory, Dartmouth College, Hanover, NH 03755, USA \\
     $^2$Center for Cosmology and Astroparticle Physics, The Ohio State University, 191 West Woodruff Lane, Columbus, Ohio 43210, USA\\
     $^3$Department of Physics and Astronomy, Johns Hopkins
     University, 3400 N.\ Charles Street, Baltimore, MD 21218, USA}

\begin{abstract}
Perhaps the most intriguing result of Planck's dust-polarization
measurements is the observation that the power in the E-mode
polarization is twice that in the B mode, as opposed to
pre-Planck expectations of roughly equal dust powers in E and B
modes.  Here we show how the E- and
B-mode powers depend on the detailed properties of the
fluctuations in the magnetized interstellar medium.
These fluctuations are classified into the slow, fast, and
Alfv\'en magnetohydrodynamic (MHD) waves, which are determined
once  the ratio $\beta$ of gas to
magnetic-field pressures is specified.  We also parametrize
models in terms of the power amplitudes and power anisotropies for
the three types of waves.  We find that
the observed EE/BB ratio (and its scale invariance) and positive
TE correlation cannot be easily explained in terms of favored
models for MHD turbulence.  The observed power-law index for
temperature/polarization fluctuations also disfavors
MHD turbulence.  We thus speculate
that the $\sim$0.1--30~pc length scales probed by these
dust-polarization measurements are not described by MHD
turbulence but, rather, probe the large-scale physics that
drives ISM turbulence.  We develop a simple
phenomenological model, based on random displacements of the
magnetized fluid, that produces EE/BB~$\simeq2$ and a
positive TE cross-correlation.
According to this model, the EE/BB and TE signals are due to
longitudinal, rather than transverse, modes in the
random-displacement field, providing, perhaps, some clue to the
mechanism that stirs the ISM.  Future investigations involving
the spatial dependence of the
EE/BB ratio, TE correlation, and local departures from
statistical isotropy in dust-polarization maps, as well as
further tests of some of the assumptions in this analysis, are
outlined.  This work may also aid in the improvement of
foreground-separation techniques for studies of cosmic microwave
background polarization.
\end{abstract}

\section{Introduction}
\label{sec:intro}

The Planck satellite has provided an extraordinary trove of
detailed information on polarized emission from dust in the
interstellar medium (ISM) of the Milky Way \citep{Ade:2014gna},
with precise power spectra measured over the multipole-moment
range $30 \lesssim \ell \lesssim 600$ \citep{Adam:2014bub}.
Since the polarization of the dust emission arises from the
alignment of spinning dust grains with the magnetic field
\citep{Chandrasekhar:1953zza,Stein:1966,Dolginov:1972,Dolginov:1976,Draine:1996nd,Draine:1996hn,Finkbeiner:2004je,Draine:2008hu,Andersson:2015},
the measurements are particularly important for the
magnetic-field structure of the ISM.

Perhaps the most surprising result from Planck is the
discovery that the E-mode power in the dust polarization is
twice the B-mode power \citep{Adam:2014bub}.  (Something similar
was noticed in WMAP, albeit with less significance, with
synchrotron polarization, \citealt{Page:2006hz}).  The
linear-polarization pattern can be decomposed geometrically into
two rotational invariants, the E (gradient) modes and B (curl) modes
\citep{Kamionkowski:1996ks,Zaldarriaga:1996xe}.  A randomly
oriented polarization map should have equal E- and B-mode
powers.  Likewise, if polarization fluctuations arise as
amplitude fluctuations with a fixed orientation, then
the E- and B-mode powers should be equal
\citep{Zaldarriaga:2001st,Kamionkowski:2014wza}.  The
state-of-the-art pre-Planck dust-polarization models
\citep{ODea:2011kx,Delabrouille:2012ye} therefore all had equal E-
and B-mode powers.  The observed
EE/BB~$\simeq2$ ratio thus comes as quite a surprise.  Planck
also finds a cross-correlation (of positive sign) between the
temperature and the E-mode component of polarization, an
empirical fact that we will also employ below.

Here we show how the observed EE/BB~$\simeq2$ ratio depends on
the detailed properties of magnetized-fluid fluctuations in the ISM.
Fluctuations in a magnetized plasma are described most generally
by the slow, fast, and Alfv\'en MHD waves; there is one for each
Fourier wavevector $\bfk$.  Models of MHD turbulence predict
the power spectra for these different types of modes as a
function of the magnitude and orientation (with respect to the
background magnetic field) of the wavevector $\bfk$
\citep{Cho:2002qn,Elmegreen:2004wj,Brandenburg:2013vya,Schekochihin:2007mw}.
A vigorous
effort, based on analytic arguments and numerical simulations,
is afoot to nail down these predictions, with much of the effort
tracing back to classic work by \citet{Iroshnikov} and
\citet{Kraichnan:1965zz} and later \citet{Shebalin:1983zz},
and more recently, for example, \citet{Goldreich:1994zz},
\citet{Lithwick:2001it}, and \citet{Cho:2002qi}.

The Planck Collaboration observed that correlations of
filamentary structures \citep{Ade:2015mbc,Adam:2014gaa} with
fluctuations in the magnetic-field orientation could account for
the observed ratio.  The Planck Collaboration made further
contact with MHD-turbulence models for the ISM in
\citet{Ade:2014hna} and \citet{Aghanim:2016uao} through
measurement of distributions of polarization magnitudes and
orientation angles.  This work does not, however, explain
how the relevant density--magnetic-field correlations arise in
terms of the fundamental modes of fluctuations in the
magnetized fluid.  There is thus room to make clearer contact
with theoretical models for a magnetized fluid.

Below we calculate the E- and B-mode amplitudes induced
by slow, fast, and Alfv\'en waves for different directions of the
background magnetic field with respect to the line of sight and
for different wavevectors $\bfk$.  Since the EE/BB~$\simeq2$
ratio seems to be relatively generic across the sky,
it must arise after averaging over all magnetic-field
orientations.  We thus then calculate the E and B
power-spectrum amplitudes, as well as the
temperature-polarization cross-correlation, obtained after
averaging over all magnetic-field and $\bfk$ orientations.  We
provide results as a function of the ratio $\beta\equiv P_g/P_H$
of the gas and magnetic-field pressures, $P_g$ and $P_H$,
respectively, and for a parameter $\lambda$ that describes the
anisotropy of the slow, fast, and Alfv\'en waves.
These calculations can then be used to assess the
validity of any particular model for MHD turbulence specified by
the power in the slow, fast, and Alfv\'en waves, and the
anisotropy of that power.

Our results suggest that for $\beta \gtrsim 1$, the observed
EE/BB ratio and temperature-polarization cross
correlation can be explained only if the power in fast waves
greatly exceeds that in slow/Alfv\'en waves, and moreover, only
if those fast waves have a nearly isotropic spectrum.  The
observations can also be explained in a low-$\beta$
(strong-field) plasma with an additional contribution from an
anisotropic spectrum of Alfv\'en waves, but only if the
slow waves are very anisotropic or somehow suppressed.  We thus
infer that the oberved EE/BB and TE are in tension with
expectations from MHD turbulence.
The apparent scale invariance of the EE/BB ratio over the
range $\ell \simeq 30-600$ and the spectral index of the
fluctuations---which disagrees with that expected from
turbulence and that seen in electron-density fluctuations on
smaller scales \citep{Armstrong:1995}---are also not easily
accommodated by current MHD-turbulence models.

We thus speculate that the $\sim$~0.1--30~pc length scales
probed by Planck may overlap the outer scale of turbulence, the
largest distance scale on which turbulence is driven.
(Alternatively, there may be new physics---e.g., associated with
the multiphase nature of the ISM
\citep{Norman:1996ba,Kritsuk:2001cm}---that is not included in
the MHD-turbulence models.)  We then develop a
simple phenomenological model, based on random displacements of
a magnetized fluid, that accounts for EE/BB~$\simeq2$ and
TE~$>0$.  We further show that the TE
correlation and large EE/BB are a consequence primarily of the
longitudinal, rather than transverse, modes in the
random-displacement field.  We surmise that this may indicate
something about the physics---perhaps stellar winds,
protostellar outflows, supernovae \citep{Lacki:2013nda,Padoan:2016}, or
Galactic spiral shocks \citep{Kim:2006ny}---that drives
small-scale turbulence in the ISM.

Directions for future related research include improved
measurement of the Planck TE cross-correlation coefficient
calculated here; studies of the variation of EE/BB and TE (that
arise from variations in the background-magnetic-field
orientation) across the sky; searches for local departures from
statistical isotropy that arise for the same reason; and more
precise measurements of the $\ell$ dependence of the dust power
spectra.  Moreover, as discussed below, we assume here that the
dust density traces the plasma density, a hypothesis that we
argue is reasonable, although one whose validity requires
further investigation.  There are thus further studies that should
be done--including the frequency dependence of the E/B/T maps,
cross-correlation with synchrotron-polarization maps, and
perhaps cross-correlation with polarized-starlight surveys---to
test further this hypothesis.
Finally, a better understanding of the physics
responsible for polarized dust emission may also aid in the
development of algorithms to separate the CMB-polarization
signal from polarized dust emission \citep{Dunkley:2008am} and
thus help advance the quest for inflationary gravitational waves
\citep{Kamionkowski:1996zd,Seljak:1996gy,Kamionkowski:2015yta}.

Such developments must not necessarily await the next flagship satellite
mission: there are prospects for considerable improvements in
dust-polarization maps on small patches of sky with suborbital
experiments \citep{Kovetz:2015pia} such as BLASTPol
\citep{Fissel:2010aa}, BFORE
\citep{Niemack:2015qta}, TOLTEC \citep{Wilson:2016}, or PILOT
\citep{Misawa:2014hka}.
Measurements of Galactic synchrotron and/or dust polarization on
larger angular scales will be improved, for example, with CLASS
\citep{Essinger-Hileman:2014pja} or LiteBird
\citep{Matsumura:2013aja}.  Analyses similar to those we discuss
can also be applied to maps of starlight
polarization \citep{Goodman:1990,Heiles:1996,Fosalba:2001wr} or
neutral-hydrogen filaments \citep{Clark:2015cpa}, although the
polarization strength is small, and the sparse sampling and the range of distances to stars complicates
the E/B mode analysis.  Moreover, similar
analyses may be employed to understand, with dust-polarization
maps, magnetic-field structure in specific molecular clouds
\citep{Pelkonen:2006fp,Kataoka:2012ci,Koch:2013rea,Soler:2013kga}.

This paper is organized as follows:  In Section \ref{sec:EB} we
review the E/B decomposition of a polarization map.  We review
the relevant properties of MHD waves in Section
\ref{sec:mhd}.  Section \ref{sec:EBmodes} calculates the E and B
amplitudes that arise from slow, fast, and Alfv\'en waves.  Section
\ref{sec:powerspectra} discusses calculation of the power
spectra.  Section \ref{sec:results} presents the results of the
calculations.  In Section \ref{sec:interpretation} we provide
some possible interpretations of the data in terms of
MHD-turbulence models and also discuss the tension with
expectations from favored MHD-turbulence models.  We therefore
consider, in Section \ref{sec:rand}, a simple phenomenological
model of random displacements in a magnetized fluid that results
in EE/BB~$\simeq2$ and TE~$>0$.  We then conclude and enumerate
several further research directions in Section
\ref{sec:conclusions}.

To avoid confusion with the E/B decomposition of polarization maps, we use $\bfH$ to denote the magnetic field. The c.g.s. system of units is used.

\section{Review of the E/B decomposition of a polarization map and projection effects}
\label{sec:EB}

Here we recall some basic properties of the decomposition of a polarization map into E and B modes, and the way in which 3-dimensional emitting structures appear on the 2-dimensional sky.  We consider a map of the linear polarization on a patch of sky
sufficiently small to be assumed flat, and of solid angle $\Omega$. We assume the emission to be optically thin, which is a good approximation at microwave frequencies.

The polarization is
specified in terms of Stokes   
parameters $Q(\bftheta)$ and $U(\bftheta)$, measured with
respect to some $\hattheta_x$-$\hattheta_y$ axes in the
plane of the sky, which can then be written as a complex
polarization $\Pi(\bftheta)=Q(\bftheta)+i U(\bftheta)$.\footnote{In the CMB literature this is often written $P(\bftheta)$, but here we write $\Pi$ to avoid confusion with the 3D power spectrum.} The
map is equivalently represented by the Fourier transform,
\begin{equation}
     \tilde \Pi(\bfell) = \int_\Omega \, d^2\bftheta\,
     \Pi(\bftheta) e^{- i \bfell \cdot \bftheta}.
\label{eqn:Fourier}     
\end{equation}
The density of Fourier modes in the 2-dimensional $\bfell$-plane is $\Omega/(2\pi)^2$.

The Stokes parameters, and the complex polarization, are not
rotational invariants; under a rotation of the coordinate axes
by an angle $\alpha$, the polarization transforms as $\Pi \to
\Pi e^{2 i \alpha}$.  The polarization field can be represented in
terms of rotational invariants $E$ and $B$. In Fourier
space these are
\begin{equation}
     (\tilde E + i \tilde B)(\bfell) = (\tilde Q+i \tilde
     U)(\bfell) e^{-2i \psi_{\bfell}}
\label{eqn:EB}
\end{equation}
\citep{Kamionkowski:1996zd,Seljak:1996gy,Kamionkowski:1996ks,Zaldarriaga:1996xe,Seljak:1996ti,Cabella:2004mk,Kamionkowski:2015yta},
where $\psi_{\bfell}$ is the angle that $\bfell$ makes with
$\hattheta_x$, i.e.\ $\tan\psi_{\bfell} = \ell_y/\ell_x$. The power spectra measured by Planck are then
$C_\ell^{\rm EE} = \langle{|\tilde E(\bfell)|^2}\rangle/\Omega$ and
$C_\ell^{\rm BB} = \langle{|\tilde B(\bfell)|^2}\rangle/\Omega$, where the
average is over all $\bfell$ of magnitude $\ell$.\footnote{The factor of $\Omega$ arises from the density of Fourier modes; the ``usual'' equation would read $\langle \tilde E^\ast(\bfell) \tilde E(\bfell') \rangle = (2\pi)^2\delta^{\rm D}(\bfell-\bfell') C_\ell^{\rm EE}$, where $\delta^{\rm D}$ is the Dirac $\delta$-function. For a density of modes $\Omega/(2\pi)^2$, we have $(2\pi)^2\delta^{\rm D}(\bfell-\bfell') \rightarrow \Omega\delta^{\rm K}_{\bfell,\bfell'}$, where $\delta^{\rm K}$ is the Kronecker $\delta$-symbol.}

The observed polarization signal $\Pi$ is typically measured in units of $\mu$K$_{\rm CMB}$, and its angular power spectra $C_\ell^{\rm EE/BB}$ have units of $\mu$K$_{\rm CMB}^2$. However, for optically thin emission, the polarization is related to the polarized emissivity $\varepsilon_\Pi$ via
\begin{equation}
\Pi(\bftheta) = \int_0^\infty \varepsilon_\Pi(r\hat{\mathbf n}(\bftheta))\,dr,
\label{eq:Pi-theta}
\end{equation}
where $\hat{\mathbf n}(\bftheta)$ is the 3-dimensional unit vector in the direction corresponding to angular position $\bftheta$. The emissivity $\varepsilon_\Pi$ (and its components, $\varepsilon_Q$ and $\varepsilon_U$) have units of $\mu$K$_{\rm CMB}$\,pc$^{-1}$, and its 3-dimensional power spectra $P_{\varepsilon,\rm EE}({\mathbf k})$ and $P_{\varepsilon,\rm BB}({\mathbf k})$ have units of $[\varepsilon_P^2]\times\,$[volume], or $\mu$K$_{\rm CMB}^2$\,pc.

For small angles or $\ell\gg 2$, the relation of 3-dimensional and 2-dimensional power spectra is usually obtained via the Limber approximation. This begins with breaking the line-of-sight integral, Eq.~(\ref{eq:Pi-theta}), into a series of boxes along the line of sight of width $\Delta r_i$. In each box, the emissivity can be Fourier-transformed to $\tilde\varepsilon_\Pi({\mathbf k})$:
\begin{equation}
\tilde\varepsilon_\Pi({\mathbf k}) =\! \int_V \! \varepsilon_\Pi({\mathbf x})\,e^{-i{\mathbf k}\cdot{\mathbf x}}\,d^3{\mathbf x}
\;\leftrightarrow\;
\varepsilon_\Pi({\mathbf x}) = \sum_{\mathbf k} \tilde\varepsilon_\Pi({\mathbf k})\,e^{i{\mathbf k}\cdot{\mathbf x}},
\end{equation}
where the Fourier wave vector ${\mathbf k}$ has (i) a transverse component ${\mathbf k}_\perp$ with a density of modes $r^2\Omega/(2\pi)^2$, and (ii) a line-of-sight component $k_\parallel = 2\pi n/\Delta r_i$ with $n\in{\mathbb Z}$. The volume of the box is $V = r^2\Omega \Delta r_i$. These transformed quantities satisfy
\begin{eqnarray}
\langle \tilde\varepsilon_E^\ast({\mathbf k}) \tilde\varepsilon_E({\mathbf k}') \rangle
&=& (2\pi)^3 \delta^{\rm D}({\mathbf k}-{\mathbf k}') P_{\varepsilon,\rm EE}({\mathbf k})
\nonumber \\
&=& r^2\Omega\Delta r_i\,\delta^{\rm K}_{{\mathbf k},{\mathbf k}'}P_{\varepsilon,\rm EE}({\mathbf k}).
\label{eq:Pow3D}
\end{eqnarray}
Only the transverse ($n=0$ or $k_\parallel=0$) modes, i.e.\ those with ${\mathbf k}$ in the plane of the sky, survive radial integration. They relate to the projected polarization via
\begin{equation}
\tilde\Pi(\bfell) = \sum_i \frac1{r^2} \tilde\varepsilon_\Pi({\mathbf k} = \bfell/r),
\label{eq:box-sum}
\end{equation}
where the $1/r^2$ comes from the transformation from
$d^2\bftheta$ to $d^2{\mathbf x}_\perp$ in the Fourier integral
(see Eq.~(\ref{eqn:Fourier})), and from Eq.~(\ref{eq:Pow3D}) the
2-dimensional power spectrum is
\begin{equation}
C_\ell^{\rm EE}
\approx \sum_i \frac{P_{\varepsilon,\rm EE}(k=\ell/r)}{r^2} \Delta r_i
\approx \int_0^{r_{\rm max}} \frac{P_{\varepsilon,\rm EE}(k=\ell/r)}{r^2}\,dr,
\label{eq:Limber}
\end{equation}
where $r_{\rm max}$ is the maximum distance from which dust
emission is seen.  Eq.~(\ref{eq:Limber}) is the Limber equation,
as commonly used in cosmology. The derivation contains two
subtle assumptions: (i) each box can be treated as a
statistically homogeneous medium; and (ii) when squaring
Eq.~(\ref{eq:box-sum}) and taking the expected value, we can
neglect correlations between different boxes $i\neq j$.

In most of this paper, we will focus our attention on the ratios of the power spectra, $P_{\varepsilon,\rm EE}(k)/P_{\varepsilon,\rm BB}(k)$, or correlation coefficients between the E-mode and temperature $r = P_{\varepsilon,\rm TE}(k)/[P_{\varepsilon,\rm TT}(k)P_{\varepsilon,\rm EE}(k)]^{1/2}$. It is easily seen from Eq.~(\ref{eq:Limber}) that the corresponding ratio in the power spectrum, $C_\ell^{\rm BB}/C_\ell^{\rm EE}$, is a suitably weighted average of $P_{\varepsilon,\rm EE}(k)/P_{\varepsilon,\rm BB}(k)$ along the line of sight. Therefore, in attempting to explain the observed EE/BB ratio, we focus on the 3-dimensional power spectrum. When we consider the scale dependence of the polarization power spectrum, we will have to return to the full version of Eq.~(\ref{eq:Limber}).

\section{Magnetohydrodynamic Waves}
\label{sec:mhd}

A compressible magnetized plasma can, in the MHD limit, carry
three different types of waves, linear combinations of the two
transverse-vector components of the magnetic field $\bfH$ (since
the requirement
$\mathbf{\nabla} \cdot \bfH=0$ removes the longitudinal-vector degree of
freedom) and the plasma-density degree of freedom.  Here we
briefly reprise the properties, relevant for this work, of these
three MHD waves, which are classified into Alfv\'en, slow, and
fast modes.

We consider a magnetized plasma at rest with a homogeneous
magnetic field $\bfH_0$ and then consider small perturbations
parametrized in terms of a magnetic-field perturbation $\delta
\bfH(\bfx,t)$ and plasma velocity $\bfv(\bfx,t)$.  In the MHD
limit, the perturbation, velocity, and background field are
related (in Fourier space) by
\begin{equation}
     \omega \delta \bfH = -\bfk \times
     (\bfv \times \bfH_0),
\label{eqn:mhd}
\end{equation}
where here $\delta\bfH$ and $\bfv$ are taken to be the
magnetic-field and velocity amplitudes of this particular
Fourier mode.

\subsection{Alfv\'en waves}

The Alfv\'en wave has a velocity perpendicular
to both $\bfk$ and $\bfH$, and it has a dispersion relation
$\omega = \pm a k \cos\alpha$, where $a= H_0 (4\pi \rho)^{-1/2}$
is the Alfv\'en speed (and $\rho$ the plasma mass density), and
$\cos\alpha = \hatk \cdot \hatH_0$. For
this wave, $\delta \bfH = \pm H_0 (\bfv/a)$.
The continuity equation, $(\partial n/\partial t) + \nabla
\cdot( n \bfv)=0$, provides a relation, $(\delta n /n_0) = \bfk
\cdot \bfv / \omega$, between the fractional density perturbation
$(\delta n/n_0)$ and the velocity.  Since $\bfk \perp \bfv$ in the
Alfv\'en wave, these waves have no associated density
perturbation. We thus write,
\begin{equation}
     \delta \bfH = -\frac{v H_0}{a} \hata,
\label{eqn:Alfvenamplitude}
\end{equation}
where $\hata \equiv \hatk \times \hatH/\sin\alpha$ is the 
unit vector perpendicular to $\bfk$ and $\bfH$.

\subsection{Slow/fast waves}

The slow and fast waves both have
magnetic-field perturbations in a direction $\hattheta =
-\hatk\times (\hatk \times \hatH)/\sin\alpha$ perpendicular to
$\hatk$ and $\hata$.  The slow wave has a displacement in 
direction $\hatxi_s \propto \cos\alpha \hatH + \zeta_s
\sin\alpha \hatk_\perp$, where $\hatk_\perp$ is a unit vector in
the $\bfk$-$\bfH$ plane perpendicular to $\hatH$, and the
fast-wave is in the orthogonal direction, $\hatxi_f \propto
\zeta_f \cos \alpha \hatH + \sin\alpha \hatk_\perp$.  Here,
\begin{eqnarray}
     \zeta_s &= &\frac{1-\sqrt{D} -\beta/2}{1+\sqrt{D} +\beta/2}
     \cot^2\alpha, \nonumber \\
     \zeta_f &= &\frac{1-\sqrt{D} +\beta/2}{1+\sqrt{D} -\beta/2}
     \tan^2\alpha,
\label{eqn:zetas}
\end{eqnarray}
where $D = (1+\beta/2)^2 - 2\beta\cos^2\alpha$, and $\beta =
P_g/P_H$ is the ratio of gas pressure to magnetic-field
pressure.  In the strong-field limit $\beta \to 0$, and
$\beta\to\infty$ in the weak-field limit.

From Eq.~(\ref{eqn:mhd}) it follows that for the slow wave,
\begin{equation}
     \delta \bfH = \frac{kv H_0}{\omega} \frac{\zeta_s
     \sin\alpha}{
     \left( \cos^2\alpha + \zeta_s^2\sin^2\alpha \right)^{1/2}}
     \hattheta, 
\label{eqn:slowdeltaH}
\end{equation}
and for the fast wave,
\begin{equation}
     \delta \bfH = \frac{kv H_0}{\omega} \frac{\sin\alpha}{
     \left( \zeta_f^2 \cos^2\alpha + \sin^2\alpha \right)^{1/2}}
     \hattheta, 
\label{eqn:fastdeltaH}
\end{equation}
where $v$ is the magnitude of the fluid velocity.  
For the
Alfv\'en wave, the relationship between the magnitudes of the
magnetic-field and velocity perturations is independent of the
orientation of $\bfk$ [cf.\ Eq.~(\ref{eqn:Alfvenamplitude})].
The same is not true, however, for the slow/fast waves.  In
addition to the explicit $\alpha$ dependence in
Eqs.~(\ref{eqn:slowdeltaH})--(\ref{eqn:fastdeltaH}), there is also an $\alpha$
dependence in $\zeta_{s,f}$ and also in the dispersion
relations,
\begin{equation}
     \left( \frac{ \omega}{k} \right)^2 = \frac{a^2}{2} (1 +\beta/2)
     \left[ 1 \pm \left( 1 - \frac{ 2 \beta
     \cos^2\alpha}{(1+\beta/2)^2} \right)^{1/2} \right],
\end{equation}
for the fast (plus sign) and slow (minus sign) waves.

The
fractional density perturbation is then found from the
continuity equation to be, for the slow wave,
\begin{equation}
     \frac{\delta n}{n_0} = \frac{kv}{\omega} \frac{\left(
     \cos^2\alpha + \zeta_s \sin^2\alpha \right)}{
     \left( \cos^2\alpha + \zeta_s^2\sin^2\alpha \right)^{1/2}},
\label{eqn:slowdeltan}
\end{equation}
and for the fast wave,
\begin{equation}
     \frac{\delta n}{n_0} = \frac{kv}{\omega} \frac{\left(
     \zeta_f\cos^2\alpha + \sin^2\alpha \right)}{
     \left( \zeta_f^2 \cos^2\alpha + \sin^2\alpha \right)^{1/2}}.
\label{eqn:fastdeltan}
\end{equation}
The final relations then are those between the
magnetic-field perturbation and the density
perturbation, and the magnetic-field perturbation and the velocity
perturbation.  They are, for the slow wave,
\begin{eqnarray}
    \frac{\delta n}{n_0} &=& \frac{|\delta\bfH|}{H_0} \left(
    \cos^2\alpha + \zeta_s\sin^2\alpha \right) /
    (\zeta_s \sin\alpha) \equiv  \frac{|\delta\bfH|}{H_0}
    g_s(\alpha), \nonumber\\
    \label{eqn:slowrelation}\\
    \frac{|\delta\bfH|}{H_0} &=& \frac{k}{\omega} \frac{\zeta_s \sin\alpha}{ \left(\cos^2\alpha +  \zeta_s^2 \sin^2\alpha \right)^{1/2}} |\mathbf{v}|\equiv |\mathbf{v}| h_s (\alpha) ,
\label{eqn:slowrelationV}
\end{eqnarray}
and for the fast wave,
\begin{eqnarray}
    \frac{\delta n}{n_0} &=&  \frac{|\delta\bfH|}{H_0} \left(
    \zeta_f \cos^2\alpha + \sin^2\alpha \right) /
    \sin\alpha \equiv \frac{|\delta\bfH|}{H_0} g_f(\alpha), \nonumber\\
    \label{eqn:fastrelation}\\
    \frac{|\delta\bfH|}{H_0} &=& \frac{k}{\omega} \frac{\sin\alpha}{ \left(\zeta_f^2 \cos^2\alpha +  \sin^2\alpha \right)^{1/2}} |\mathbf{v}|
    \equiv |\mathbf{v}| h_f(\alpha)  .
\label{eqn:fastrelationV}
\end{eqnarray}
In the case of the Alfv\'en wave, as seen in
Eq.~(\ref{eqn:Alfvenamplitude}), we have ${|\delta\bfH|}/{H_0}
= |\mathbf{v}|/a \equiv |\mathbf{v}| h_a$.  These relations allow
us to determine the E- and B-mode powers under different
assumptions about the power spectra for the different MHD waves.

\section{E and B modes induced by the slow, fast, and Alfv\'en Waves}
\label{sec:EBmodes}

\subsection{E and B amplitudes from a single Fourier mode}

\begin{figure}[htbp]
\includegraphics[scale=0.5]{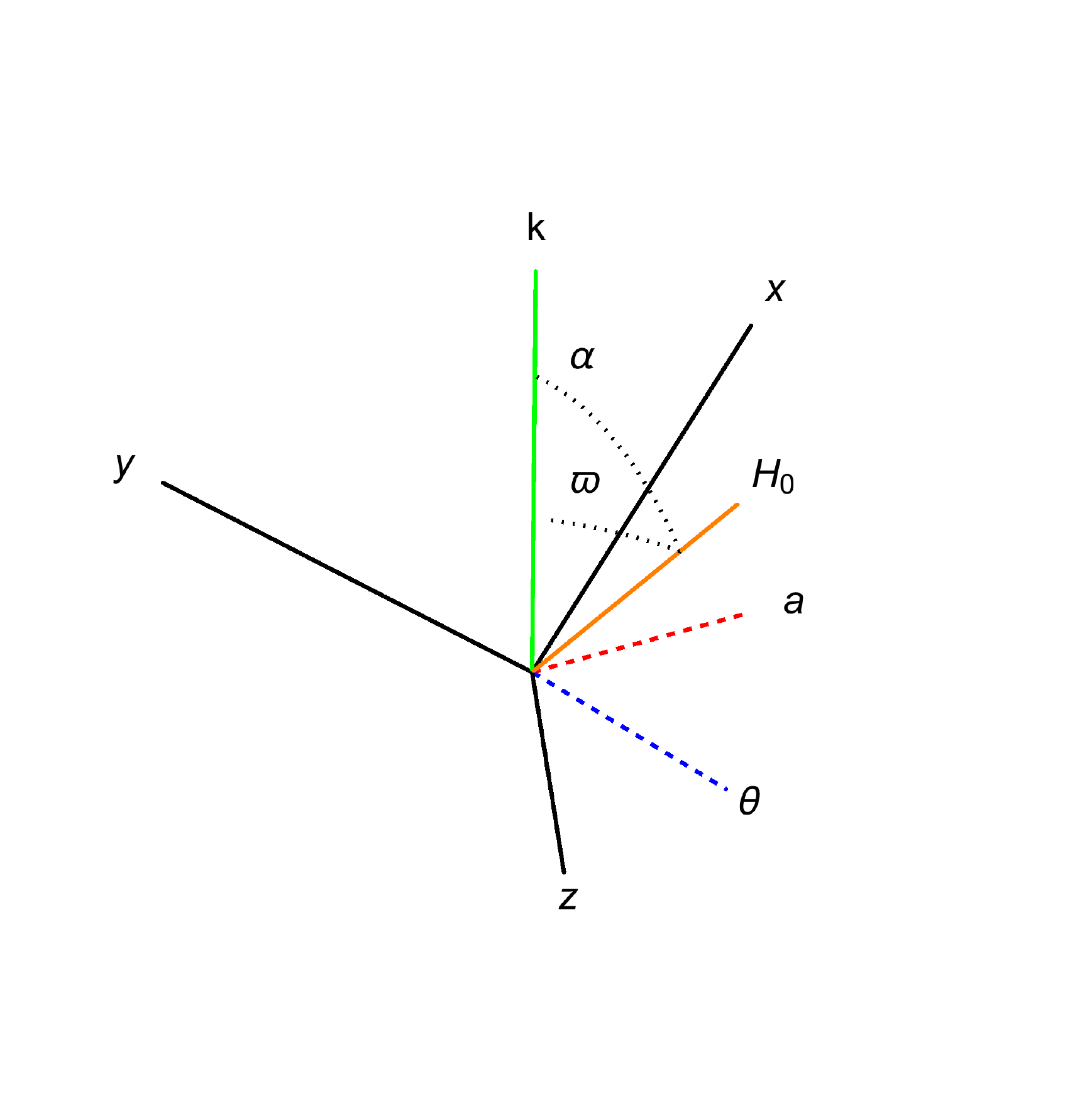}
\caption{The coordinate axes and relevant vectors are shown. The
     locations of $\mathbf{H}_0$ and $\mathbf{k}$ are shown by
     the orange and green lines, and $\mathbf{a}$ and
     $\hat{\theta}$ by the dashed red and blue lines,
     respectively. The coordinates $\alpha$ and $\varpi$
     relative to $\mathbf{k}$ are also indicated. Recall that
     the line of sight is along the $z$ direction.\\} 
\label{fig:coords}
\end{figure}

Take the line of sight to be along the $z$ axis and 
the background field $\bfH_0 = H_0 (\sin\theta,0,\cos\theta)$
in the $x$-$z$ plane at an angle $\theta$ from the line of
sight.  Consider a perturbation of
wavevector $\bfk = k \hatk = k(\cos\psi,\sin\psi,0)$ in the
$x$-$y$ plane of the sky (as the two-dimensional projections of
other modes will experience a Limber suppression)
oriented at an angle $\psi$ with respect to the $x$ axis.  The
angle $\alpha$ between $\bfk$ and $\bfH$ is then given by
$\cos\alpha = \sin\theta \cos\psi$, as illustrated in Fig.~\ref{fig:coords}.

We observe a two-dimensional projection of an emitting volume,
and the polarized emission is assumed to have the form
\begin{equation}
\varepsilon_P = \varepsilon_Q + i \varepsilon_U = A n H^\gamma(H_x+i H_y)^2,
\label{eqn:polarization}
\end{equation}
where $\gamma$ is an exponent which is equal to $-2$ if the dust
alignment is independent of the magnetic-field strength, $n$
is proportional to the dust density (and has a constant
background value $n_0$).  Here $A<0$ is a constant; its value is
taken to be negative so that the polarization is perpendicular
to the magnetic field \citep{Chandrasekhar:1953zza}.  The
sign of $A$ will be significant for the temperature-polarization
cross-correlation below.  The polarization fluctuations are
\begin{widetext}
\begin{equation}
     \delta\varepsilon_P = A n_0 H_0^{2+\gamma} \left[2 \sin\theta \,\frac{\delta
     H_x+ i \delta H_y}{H_0}
     + \gamma \sin^2\theta \,\frac{\delta H}{H_0} + \sin^2\theta \,\frac{\delta
     n}{n_0} \right],
\end{equation}
where here $\delta H = \sin\theta \delta H_x+ \cos\theta \delta
H_z$.  For a given Fourier mode of wavevector $\bfk$ transverse to the line of sight in a box of radial width $\Delta r$, the $E$
and $B$ modes will appear in wave vector $\bfell = \bfk r$, and will have the form:
\begin{equation}
     \tilde E + i \tilde B = A n_0 H_0^{2+\gamma} \frac{\Delta r}{r^2}\,e^{-2i\psi} \left[2
     \sin\theta \,\frac{\delta\tilde H_x + i \delta\tilde H_y}{H_0} + \gamma
     \sin^2\theta \,\frac{\delta \tilde H}{H_0} + \sin^2 \theta \,\frac{\delta\tilde n}{n_0} \right],
\end{equation}
which can be decomposed into
\begin{eqnarray}
     \tilde E &=& A n_0 H_0^{2+\gamma} \frac{\Delta r}{r^2}\, \left[ 2 \sin\theta \cos2\psi
     \frac{\delta \tilde H_x }{ H_0 } + 2\sin\theta \sin 2\psi \,\frac{\delta \tilde H_y}{H_0} +
     \gamma \sin^2\theta \cos 2\psi \,\frac{\delta \tilde H}{H_0} + 
     \sin^2\theta \cos 2\psi \,\frac{\delta\tilde n}{n_0} \right], \nonumber
     \\
     \label{eqn:Egen}      \\
     \tilde B &=& A n_0 H_0^{2+\gamma} \frac{\Delta r}{r^2}\, \left[ -2 \sin\theta \sin 2\psi
     \,\frac{\delta\tilde H_x }{ H_0}+ 2\sin\theta \cos 2\psi \,\frac{\delta\tilde H_y}{H_0} -
     \gamma \sin^2\theta \sin 2\psi \,\frac{\delta\tilde H}{H_0} -
     \sin^2\theta \sin 2\psi \,\frac{\delta\tilde n}{n_0} \right]. \nonumber\\
     \label{eqn:Bgen}
\end{eqnarray}

We now re-write the magnetic-field perturbations in terms of the
two transverse-vector modes, those in the $\hata$ (the Alfv\'en
wave) and $\hattheta$ (the slow and fast waves) directions:
\begin{eqnarray}
     (\delta \tilde \bfH)_a &=& \hata \delta \tilde H_a = \frac{\hatk \times
     \hatH}{\sin \alpha} \delta \tilde H_a = \frac{ (\sin\psi\cos\theta,
     -\cos\psi \cos\theta, -\sin\psi\sin\theta)}{\sin \alpha}
     \delta\tilde H_a,\nonumber \\
     (\delta\tilde \bfH)_p &=& \hattheta \delta \tilde H_p = -\frac{\hatk \times
     (\hatk \times \hatH)}{\sin\alpha} \delta \tilde H_p = \frac{
     (\sin\theta \sin^2\psi,-\sin\theta
     \sin\psi\cos\psi,\cos\theta)}{\sin\alpha} \delta \tilde H_p,
\end{eqnarray}
where $\delta \tilde H_a$ (`a' for Alfv\'en) and $\delta \tilde H_p$ (`p' for
pseudo-Alfv\'en) are the magnetic-field amplitudes for the two
modes.  These then translate to $E$ and $B$ modes,
\begin{eqnarray}
     \tilde E &=& A n_0H_0^{2+\gamma} \frac{\Delta r}{r^2} \left[ -\sin2\theta
     \frac{\sin\psi}{\sin\alpha} \frac{\delta \tilde H_a}{H_0} + \frac{ \sin^2\theta
     \left[ -2\sin^2\psi (1+\gamma \sin^2\alpha) + \gamma
     \sin^2\alpha \right]}{\sin\alpha} \frac{\delta\tilde  H_p}{H_0} + \sin^2\theta
     \cos2\psi \frac{\delta \tilde n}{n_0}
     \right],     \label{eqn:Eintermediate} \\
     \tilde B &=& An_0 H_0^{2+\gamma} \frac{\Delta r}{r^2} \left[ -\sin2\theta \frac{\cos
     \psi}{\sin\alpha} \frac{\delta\tilde H_a}{H_0} - \frac{2 \sin^2\theta \sin\psi
     \cos\psi (1+\gamma \sin^2\alpha)}{\sin \alpha} \frac{\delta\tilde H_p}{H_0} -
     \sin^2\theta \sin 2\psi \frac{\delta \tilde n}{n_0} \right].
\label{eqn:Bintermediate}     
\end{eqnarray}

For Alfv\'en waves, which have no associated density perturbation,
we are already done.  However, the fast and slow waves both have
a density perturbation.  The final step is thus to re-write the
$p$ and $n$ modes in terms of slow (`s') and fast (`f') modes
using Eqs.~(\ref{eqn:slowrelation}) and (\ref{eqn:fastrelation}).
We then obtain
\begin{eqnarray}
     \tilde E &=& A n_0H_0^{2+\gamma} \frac{\Delta r}{r^2} \left[ -\sin2\theta
     \frac{\sin\psi}{\sin\alpha} \frac{\delta\tilde  H_a}{H_0}
     + \sum_{i=s,f} \frac{\delta \tilde  H_i}{H_0} \sin^2\theta \left( \frac{
     \left[ -2\sin^2\psi (1+\gamma \sin^2\alpha) + \gamma
     \sin^2\alpha \right]}{\sin\alpha} 
     +g_i(\alpha) \cos2\psi \right)\right] \nonumber \\
     & \equiv & A n_0 H_0^{2+\gamma} \frac{\Delta r}{r^2} \sum_{i=a,s,f} f_i^{\rm
     E}(\theta,\psi) \frac{\delta \tilde  H_i}{H_0} = A n_0
     H_0^{2+\gamma} \frac{\Delta r}{r^2} \sum_{i=a,s,f} f_i^{\rm E}(\theta,\psi)
     h_i(\theta,\psi) |\mathbf{v}_i|, \\
     \tilde B &=& An_0 H_0^{2+\gamma} \frac{\Delta r}{r^2} \left[ -\sin2\theta \frac{\cos
     \psi}{\sin\alpha} \frac{\delta \tilde  H_a}{H_0}
     - \sum_{i=s,f} \frac{\delta \tilde  H_i}{H_0} \sin^2\theta \sin 2\psi
     \left(\frac{(1+\gamma \sin^2\alpha)}{\sin \alpha}
          + g_i(\alpha) \right) \right] \nonumber\\
     & \equiv & A n_0 H_0^{2+\gamma}\frac{\Delta r}{r^2} \sum_{i=a,s,f} f_i^{\rm
     B}(\theta,\psi) \frac{\delta \tilde  H_i}{H_0} = A n_0
     H_0^{2+\gamma}\frac{\Delta r}{r^2} \sum_{i=a,s,f} f_i^{\rm B}(\theta,\psi)
     h_i(\theta,\psi) |\mathbf{v}_i|.
\label{eqn:EandB}
\end{eqnarray}
\end{widetext}
The intermediate lines define the angular functions $f_i^{\rm
E,B}(\theta,\psi)$ which relate the polarization pattern to the
magnetic field fluctuations, and the conversion into velocity
fluctuations follows from
Eqs.~(\ref{eqn:slowrelationV})--(\ref{eqn:fastrelationV}).
 
\subsection{Temperature fluctuations}

The brightness temperature of the dust (synchrotron) emission is
also provided, as a function of position on the sky, by Planck
\citep{Adam:2014bub,Ade:2014zja} (WMAP, \citealt{Page:2006hz}).
Since the brightness temperature of dust emission is proportional to the dust
density, temperature fluctuations arise from fluctuations $\delta
n$ in the dust density.  The fractional intensity or temperature perturbation
is thus,
\begin{equation}
     \frac{\delta \epsilon_T}{\bar\epsilon_T} = c \frac{\delta n}{n_0},
\end{equation}
and projected through a box of width $\Delta r$ we have
\begin{equation}
\tilde T(\bfell) = c \bar\epsilon_T \frac{\Delta r}{r^2}\,\frac{\delta\tilde n(\bfk)}{n_0}.
\end{equation}
We expect $c=1$ for thermal dust emission since the physical temperature of the dust grains does not depend on the gas density (it is set by radiative equilibrium). Other dust emission mechanisms, e.g.\ spinning dust, may depend in a complicated way on the local gas density \citep[e.g.][]{Draine:1997tb,AliHaimoud:2008dc} and hence for these we may have $c\neq 1$. Note however that our focus is on the TE cross-correlation coefficient, where $c$ cancels out.

Written in terms of the wave modes, we find
\begin{equation}
     \tilde T(\bfell) = c \bar\epsilon_T \frac{\Delta r}{r^2}\, \frac{\delta\tilde  n}{n_0} = c \bar\epsilon_T \frac{\Delta r}{r^2}\, \sum_{i=s,f}
     g_i(\alpha) h_i(\alpha) |\mathbf{v}_i|.
     \label{eq:DTgh}
\end{equation}
Note that the Alfv\'en modes do not yield any density perturbations, and hence do not contribute to $\tilde T$.

\section{Calculations of Power Spectra}
\label{sec:powerspectra}

We now calculate the power in E and B modes
contributed by the three different types of waves.  Strictly
speaking, we calculate the contribution to the E- and B-mode
powers at a given 3d wavenumber $k$.  The observed 2d E- and
B-mode powers, as a function of multipole $\ell$, are then
obtained from the Limber equation which sums the contributions
of wavenumbers $k=\ell/r$, from a range of distances $r$, to a given
$\ell$.  If, however, the EE/BB ratio is scale-independent (as
we assume here and as is consistent with the
measurements), then the EE/BB ratio we calculate will also be
that in the observed 2d power spectrum.  Similar remarks apply
to the TE correlation.

\subsection{Parametrization of power anisotropies in the MHD
waves}

Since the background magnetic field
$\bfH_0$ provides a preferred direction, the
power spectra for the three types of MHD waves are not expected
to be isotropic, but should, rather, have some
$\cos\alpha$ dependence \citep{Shebalin:1983zz,Goldreich:1994zz}.
Here we parametrize the anisotropy as
\begin{equation}
     P_i(k,\cos\alpha) \equiv \VEV{ \left | \frac{\delta
     \bfH_i}{H_0} \right|_{\bfk}^2} = P_i(k) \left[h_i(\alpha)
     \right]^2 F_\lambda(\cos\alpha),
\label{eqn:powerspectra}
\end{equation}     
with
\begin{equation}
    F_\lambda(\mu)  =\begin{cases}  (\mu^2)^\lambda, & \text{if} \quad
    \lambda \geq 0, \\
    (1-\mu^2)^{-\lambda}, & \text{if} \quad \lambda \leq0.
    \end{cases}
\label{eqn:flambda}
\end{equation}
We work with power spectra for the
magnetic-field amplitudes, but have then defined, by virtue of
the $h_{s,f}(\alpha)$ in Eq.~(\ref{eqn:powerspectra}), the
anisotropy $F_\lambda(\mu)$ relative to the velocity-perturbation
amplitude.  We do so to make contact with the MHD literature,
wherein wave amplitudes are usually specified in terms of
the velocity.  With our parametrization, for $\lambda=0$ the
velocity power is isotropic; for $\lambda >0$ it is weighted
in modes of wavevector $\bfk$ parallel to $\bfH_0$; and for
$\lambda<0$, the velocity power is weighted in modes
perpendicular to $\bfH_0$.

\subsection{The EE/BB ratio}

Given that the EE/BB ratio seems to be roughly 2
everywhere on the sky, any MHD explanation of the EE/BB ratio
must provide this ratio after averaging over all magnetic-field
orientations, rather than rely on a specific orientation.  There
is also evidence that the angular average is warranted even
along an individual line of sight:  If the field direction were
exactly constant along a given line of sight, then we would
expect the fractional polarization for synchrotron radiation to be $\sim75\%$
\citep{Rybicki:1979}.  Planck obtains significantly lower values
(see, e.g., Fig.~22 in \citealt{Ade:2015qkp}), suggesting a large
dispersion in field direction even on a single line of sight.

We therefore calculate the ratios $R$ of the angle-averaged E-mode
and B-mode powers, induced by Alfv\'en, slow, and fast waves as
a function of $\beta$ and the anisotropy parameter $\lambda$.
The desired ratio is obtained
from
\begin{equation}
     R_i(\beta,\lambda) = \frac{ \int d\Omega \left[
     f_i^E(\theta,\psi) h_i(\alpha) \right]^2
     F_\lambda(\cos\alpha)} { \int d\Omega
     \left[ f_i^B(\theta,\psi) h_i(\alpha) \right]^2
     F_\lambda(\cos\alpha)},
     \label{eqn:RVformula}
\end{equation}
for $i=\{a,s,f\}$.
Evaluation of the angular averages can be simplified by
transforming to new angular coordinates $\alpha$ and $\varpi$,
through $\cos\theta = \sin\alpha \cos\varpi$,
$\sin\theta\sin\psi = \sin\alpha\sin\varpi$, and $\sin\theta
\cos\psi = \cos\alpha$.  These then are polar coordinates for
the location of $\mathbf{H}_0$ about the $k$ axis, rather than
the $z$ axis, as seen in Fig.~\ref{fig:coords}.  We then
integrate over $d\Omega=\sin\alpha \, d\alpha\, d\varpi$.

\subsection{The temperature-polarization cross-correlation}

Temperature fluctuations will arise from fluctuations in the
density field, in accordance with Eq.~(\ref{eq:DTgh}).
The Alfv\'en modes do not contribute to temperature
fluctuations. The slow and fast modes, however, should set up a
correlation between the temperature and E-mode polarization. (The
TB and EB cross-correlations vanish after averaging over
angles.)  The relative amplitudes of the polarization and
temperature fluctuations depend on a polarization fraction and
the constant $c$, and so we work instead with a
cross-correlation coefficient,   
\begin{widetext}
\begin{equation}
     r_i(\lambda) =   \frac{ \int d\Omega \left[ g_i(\alpha)
     h_i(\alpha)\right] \left[ f_i^E(\theta,\psi) h_i(\alpha)
     \right] F_\lambda(\cos\alpha)}  {\sqrt{ \int d\Omega \left[
     g_i(\alpha) h_i(\alpha)\right]^2 F_\lambda(\cos\alpha)}
     \sqrt{ \int d\Omega \left[ f_i^E(\theta,\psi)
     h_i(\alpha)\right]^2 F_\lambda(\cos\alpha)} },
\end{equation}
\end{widetext}
which corresponds to the ratio $TE/\sqrt{(TT)\, (EE)}$.

\section{Results}
\label{sec:results}

\begin{figure*}[htbp]
\begin{centering}
\includegraphics[width=7in]{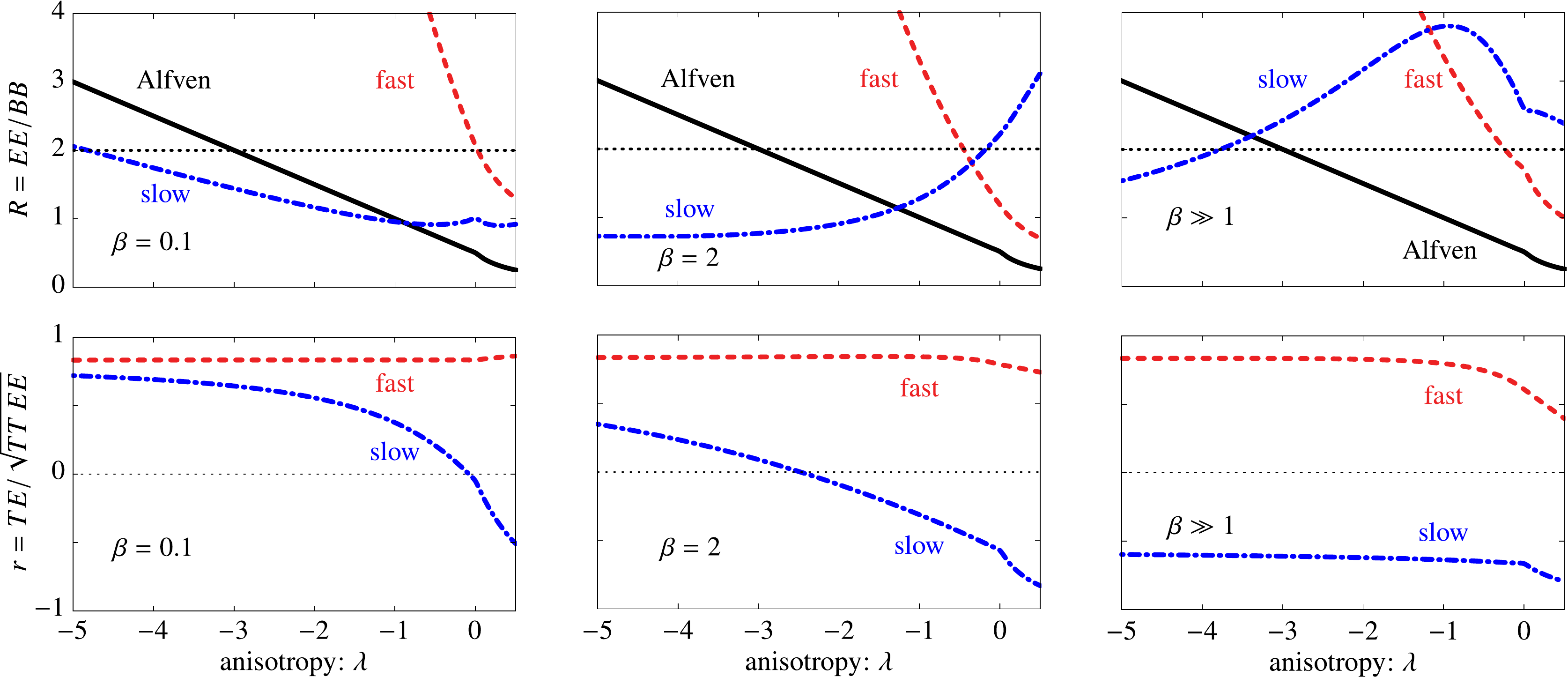}
\caption{The EE/BB ratio and cross-correlation coefficient are
     shown as a function of the velocity power spectrum
     anisotropy index $\lambda$ for $\beta=0.1,\,2,$ and $\beta
     \gg 1$. The solid (black), long
     dashed (red), and dot-dashed (blue) curves are for Alfv\'en,
     fast, and slow magnetosonic waves, respectively. The
     observed EE/BB ratio is indicated by the thin dashed
     (black) line in the upper panels. The positive cross correlation TE
     is indicated by the thin dotted (black) line in the lower panels. } 
\label{fig:panels}
\end{centering}
\end{figure*}

The EE/BB ratio and cross-correlation coefficients are shown in
Fig.~\ref{fig:panels} for a strong magnetic field ($\beta=0.1$),
equipartition ($\beta=2$), and weak field ($\beta \gg 1$). 
The two observational constraints, EE/BB~$\simeq2$ and TE~$>0$,
can be satisfied by a nearly
isotropic fast mode, for a wide range of $\beta$, or by a
strongly anisotropic slow mode, with $\beta \lesssim 2$. More
specifically, for a fast wave with $\beta=0.1$, an isotropic
spectrum ($\lambda=0$) gives EE/BB~$\simeq 2$ and
cross-correlation coefficient $r\simeq0.8$. For a slow wave with
$\beta=0.1$, too, a strongly anisotropic spectrum with
$\lambda\sim -5$ gives EE/BB~$\simeq2$ and cross-correlation
coefficient $r\simeq0.7$. The constraints cannot be satisfied by a
pure Alfv\'en wave, since this incompressible mode creates no
intensity fluctuation and therefore no cross correlation. 

All the results illustrated assume $\gamma=-2$. However, we have
also examined cases in which the polarization amplitude is correlated
with the magnetic field, $\gamma > -2$, as well as the inverse
case, $\gamma < -2$.  We find that our results for the EE/BB
ratio and TE cross correlation are not strongly sensitive to the
dust-alignment index in the range $-5/2 < \gamma < -3/2$.

\section{Interpretations}
\label{sec:interpretation}

In this Section we try to make sense of the observations within
the context of models for the ISM.  We first consider
MHD-turbulence models and conclude that they are unlikely to
provide the whole story.  We then speculate that the Planck
dust-polarization data may alternatively reflect the physics
driving turbulence and/or involve new physics beyond that
included in the MHD-turbulence models we consider here.

\subsection{MHD turbulence?}

\subsubsection{EE/BB ratio and TE correlation}

There are some important qualitative
conclusions about MHD-turbulence models that can be inferred
from the observations EE/BB~$\simeq2$ and TE~$>0$.  (Strictly speaking, the
cross-correlation coefficient we calculate here has not yet been
provided by Planck.  We estimate it by comparing Figs.~2 
and B1 in \citet{Adam:2014bub} with Fig.~D1 in
\citet{Ade:2014zja}.  There are uncertainties here: the
cuts and assumptions that went into the latter figure are not
necessarily as those that went into the first two.  Even so, we
infer that the cross-correlation coefficient is reasonably
large and, more importantly, positive.)  The models generally
predict \citep{Cho:2002qi} that: (a) slow/Alfv\'en waves should
have similar power spectra; (b) the slow/Alfv\'en should
preferentially populate modes perpendicular to the magnetic
field ($\lambda<0$ in our parlance); (c) the fast modes should
be largely uncoupled from the slow/Alfv\'en modes; and (d) the
fast modes should be nearly isotropic ($\lambda \simeq 0$).

\begin{figure*}[htbp]
\begin{centering}
\includegraphics[scale=0.45]{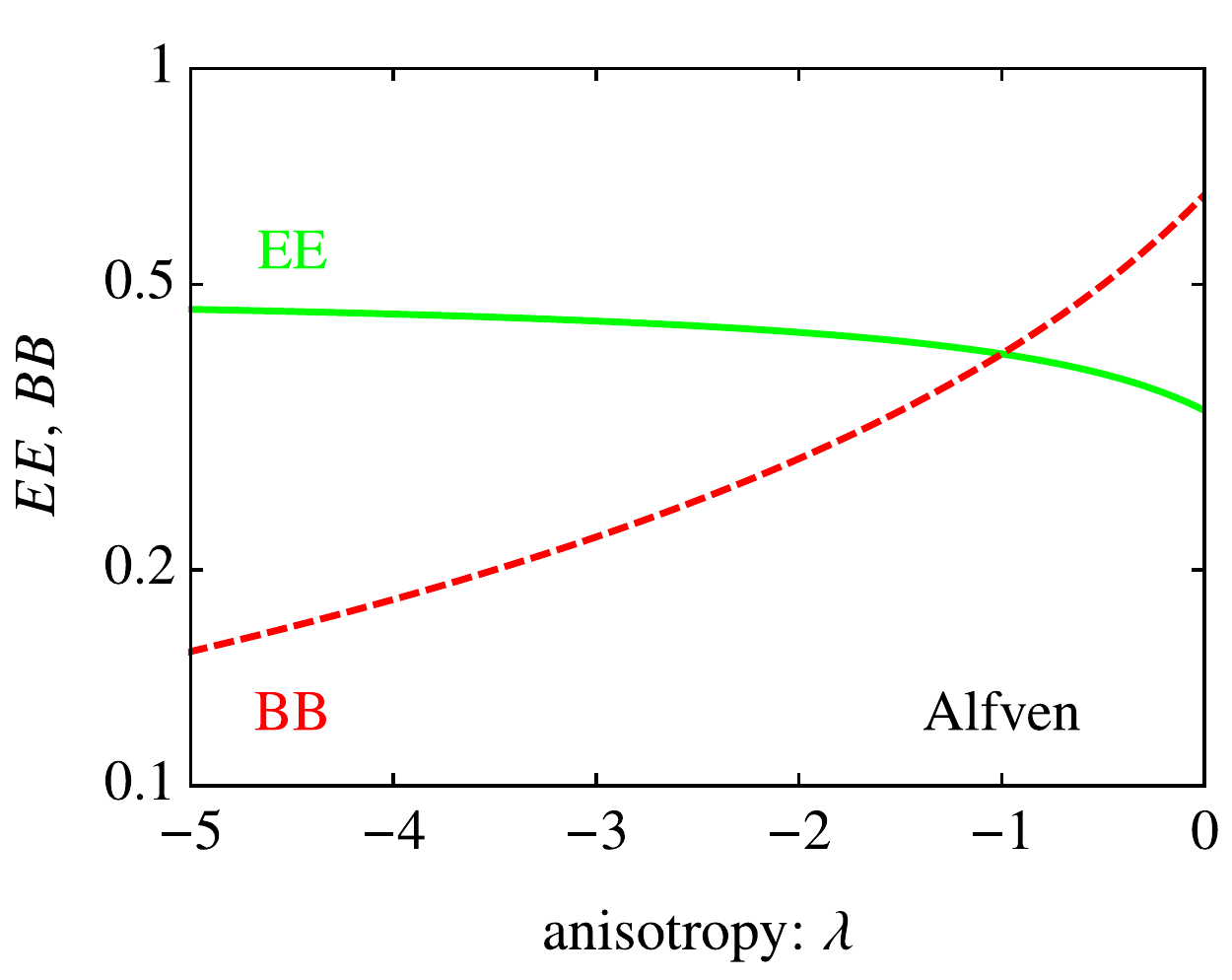}
\includegraphics[scale=0.45]{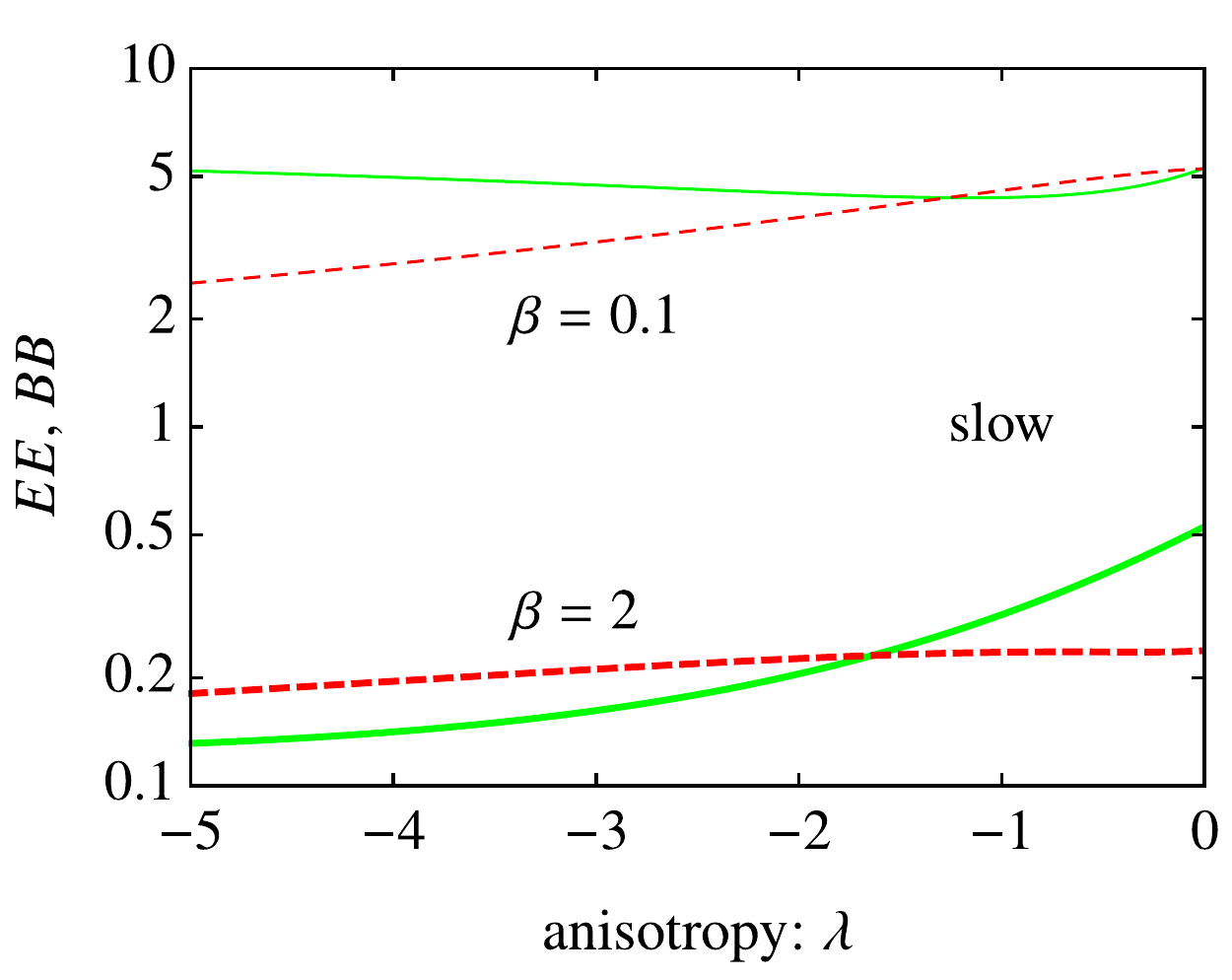}
\includegraphics[scale=0.45]{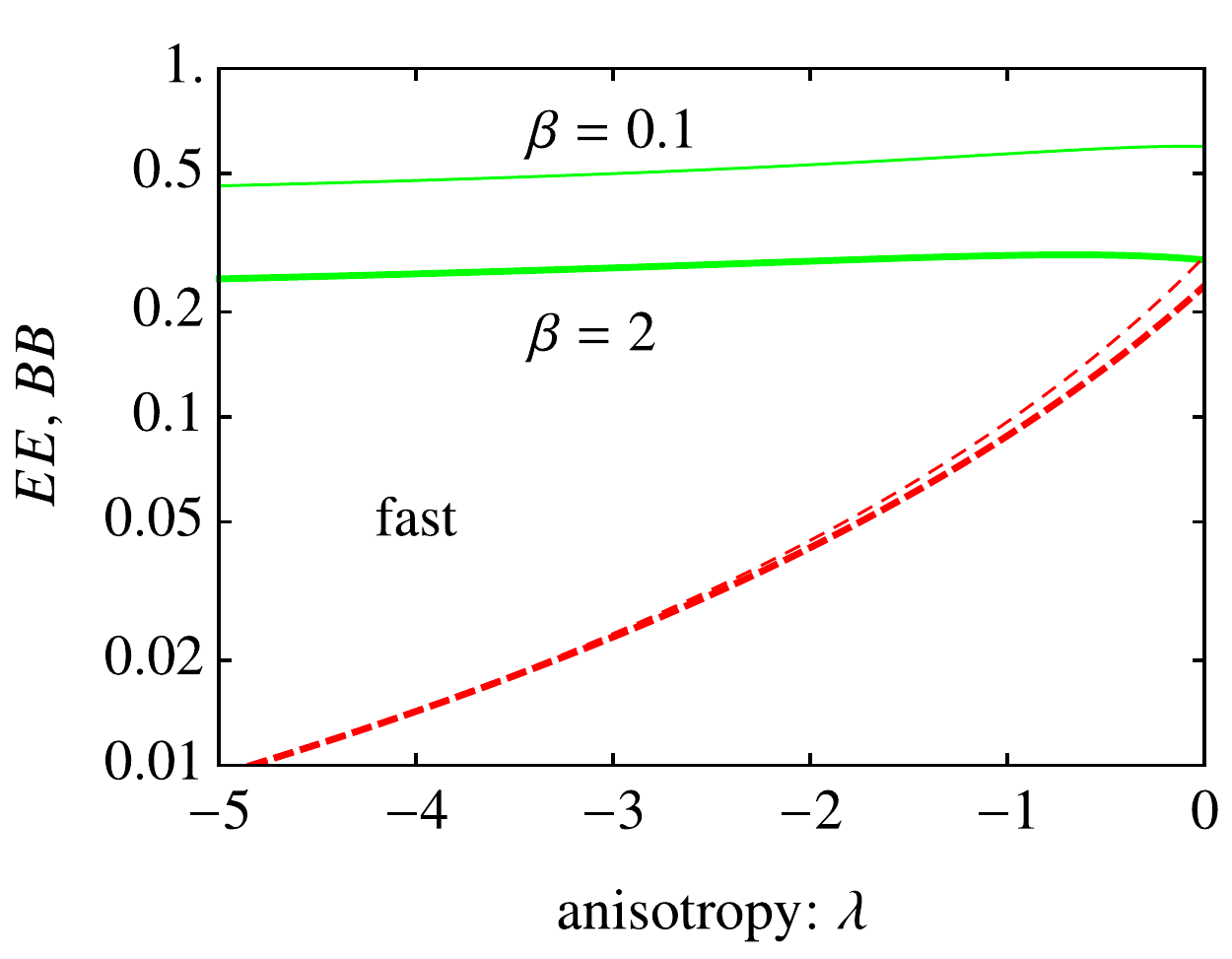}
\caption{The power EE and BB, normalized to the power in
     velocity fluctuations, are shown for each of the
     velocity modes, for the representative cases $\beta=2$ (thick lines) and $\beta=0.1$ (thin lines). As
     labeled in the figures, solid (green) are EE and dashed (red) are BB.} 
\label{fig:EBpoweq}
\end{centering}
\end{figure*}

We also need to consider the total E- and B-mode polarization
powers contributed, for fixed angle-averaged
velocity-perturbation power, by each of the different types of
MHD waves.  These are plotted in Fig.~\ref{fig:EBpoweq} for
$\beta=0.1$ and $\beta=2$ (the results for $\beta \gg 1$ are
similar to those for $\beta=2$).  For $\beta\gtrsim 1$, the
polarization powers contributed by all three types of waves are
roughly similar.  However, the polarization power in slow modes
scales inversely with $\beta$ as $\beta \to 0$.  Physically,
this occurs because $(\omega/k)\to 0$ in this limit, indicating
a vanishing restoring force.  The fluid displacements, and
thus density perturbations, become large.  Thus, the EE/BB ratio
and TE correlation will receive disproportionately large
contributions from slow modes in a low-$\beta$ plasma.

Looking at Fig.~\ref{fig:panels}, along with Fig.~\ref{fig:EBpoweq},
we see that the combination of
the two constraints (EE/BB~$\simeq2$ and TE~$>0$) very seriously
restricts the range of allowable models.  There seem to be two
possibilities:  (1) A nearly isotropic spectrum of fast waves
provides positive cross-correlation and EE/BB~$\simeq2$ for any
$\beta$.  A combination of slow/Alfv\'en waves is disallowed, on
the other hand for $\beta\gtrsim1$.  Thus, the observations can
be explained if $\beta \gtrsim1$ and Alfv\'en/slow waves are
somehow suppressed.  (2) For $\beta \ll1$, Alfv\'en waves can
produce EE/BB~$\simeq2$ if sufficiently anisotropic, but they
contribute nothing to TE. Slow modes can, if sufficiently
anisotropic, also contribute EE/BB~$\simeq2$ and a positive TE.
Given the theoretical expectation that the velocity power in
slow and Alfv\'en waves is comparable, the slow waves will
dominate at low $\beta$, and thus the anisotropy must be even
greater to account for the observations.

The fettle of either of these MHD-turbulence interpretations is
damaged by the relative uniformity---as best can be
determined---of the EE/BB ratio and TE correlation across the
sky.  The ISM is a complicated system that is likely to display
considerable variation in the parameters $\beta$ and $\lambda$
and the relative contributions of strong/fast/Alfv\'en waves.
While there are indeed pockets of the MHD-turbulence parameter
space that can account for the observed EE/BB and TE, these
predictions will not be robust if there is considerable
variation of $\beta$, $\lambda$, or the mix of slow/fast/Alfv\'en
waves within the ISM.

\subsubsection{Scale-dependent anisotropy?}

The observed power-law indexes for the $\ell$
dependences of the EE and BB power spectra agree to roughly a
percent and are also very similar to those for the TT and TE
power spectra \citep{Adam:2014bub}.  As the
Figures indicate, the EE/BB ratios can depend quite a bit on the
anistropy parameter $\lambda$.  Thus, if the power anisotropy is
scale-dependent, as expected in MHD turbulence
\citep{Goldreich:1994zz,Cho:2002qi}, then one might expect to see different
power-law indexes for E modes and B modes.  Some caution should
be used in drawing this conclusion since a given multipole
moment $\ell$ receives contributions from emission at a variety
of line-of-sight distances $r$, and thus a variety of
wavenumbers $k \sim \ell/r$.  Still, we infer that there is no
dramatic variation of the MHD power anisotropy with over the
$\sim 0.1-30$~pc length scales probed by Planck.

\subsubsection{The wavenumber scaling}
\label{sec:wavenumberscaling}

There is also a disparity between the spectral index $\nu\simeq 2.4$
measured for the TE/EE/BB/TT power spectra, $C_\ell \propto
\ell^{-\nu}$, and that, $\kappa \simeq 3.67$, in the
three-dimensional power spectrum, $P(k)\propto k^{-\kappa}$
expected in MHD turbulence.  The two exponents are related
through the Limber equation, Eq.~(\ref{eq:Limber}).  If
the three-dimensional
power spectrum is well-approximated by a single power law over
the relevant distance scales, then the two-dimensional power
spectrum $C_\ell$ will also be a power law and, moreover, with
the same spectral index, $\nu=\kappa$.  Given that the maximum
distance from which we see dust emission (at least at high
Galactic latitudes) is $r_{\rm max} \simeq 100-200$~pc, the
range of physical length  scales probed by Planck measurements
over $\ell\simeq 30-600$ is roughly $L\sim 0.1-30$~pc, where
$L=2 \pi/k$.

\subsection{An outer scale?}

Turbulence is expected, however, to be described by a power law
only below some outer distance scale
$L$, or for wavenumber $k\gtrsim k_c \sim 2\pi L^{-1}$.  Suppose, for
example, that the power is $P(k)=0$ for $k<k_c$ and $P(k)
\propto k^{-\kappa}$ for $k>k_c$ (and with $q(r)=$~constant).
In this case, we expect $C_\ell \propto \ell^{-1}$ for
$\ell \ll \ell_c \equiv k_c r_{\rm max}$ and $C_\ell \propto
\ell^{-\kappa}$ for $\ell \gg \ell_c$.  It is conceivable
that the apparent power-law index $\nu=2.42$ approximates the
scaling if the $\ell=30-600$ range over which the
measurements are done contains the characteristic multipole
$\ell_c$ that separates the $C_\ell \propto
\ell^{-1}$ low-$\ell$ behavior to the $C_\ell \propto
\ell^{-\kappa}\sim \ell^{-3.67}$ behavior at higher $\ell$.  If so,
then the outer scale is (taking $\ell_c\simeq100$ and $r_{\rm
max} \sim100$~pc) $L\sim 10$~pc, a reasonable value
and not too different from the $\sim$~pc outer
scale inferred from Faraday rotation and depolarization of
extragalactic radio sources \citep{Haverkorn:2008tb}.
If the $\ell=30-600$ range does indeed correspond
to the outer scale of turbulence, then guidance from
MHD-turbulence modeling about the power in slow/fast/Alfv\'en
waves may be inappropriate.  The observations may then have more
to do with the large-scale physics---for example, stellar winds,
protostellar outflows, supernovae
\citep{Lacki:2013nda,Padoan:2016}, or Galactic spiral 
shocks \citep{Kim:2006ny}---driving the turbulence, rather than the 
turbulence itself.  In this case, the power-law behavior in
$C_\ell$ should be only an approximation, and it should be
found, with improved measurement, to be shallower at lower
$\ell$ and steeper at higher $\ell$.

If this interpretation is correct, then extrapolations of
foreground power based on measurements at $30 \lesssim \ell
\lesssim 600$ to lower $\ell$ may be overestimating the
low-$\ell$ CMB foregrounds.  If so, this will be good news
\citep{Kamionkowski:2015yta} for experiments, such as
CLASS \citep{Essinger-Hileman:2014pja} and LiteBird
\citep{Matsumura:2013aja}, that go to low $\ell$ to seek this
signal.

\subsection{Warm/neutral transition?}

Another possibility is that the ISM is not described by the
conventional MHD-turbulence models.  For example, it is well
known that the interstellar medium is a multi-phase medium.  If
there is some instability that allows transitions, for example,
between a warm neutral phase and a cold neutral phase
then the ISM equation of state may be more
complicated than that assumed in the standard MHD analysis
\citep{Norman:1996ba,Kritsuk:2001cm}.  If
so, then the normal modes of the system may not necessarily
correspond to the standard slow/fast/Alfv\'en waves---for any value
of $\beta$---but rather consist of some other linear
combinations of them.

\subsection{Does dust trace plasma?}

The MHD approximation assumed here requires the magnetic-field
lines to be tied to the plasma, and the relations
[Eqs.~(\ref{eqn:slowrelation}) and (\ref{eqn:fastrelation})]
derived above are between the magnetic-field and plasma-density
perturbations.  Strictly speaking, though, the quantity $\delta
n$ is the perturbation to the {\it dust} density.  In deriving
Eqs.~(\ref{eqn:slowrelation}) and (\ref{eqn:fastrelation}), we
have assumed that the dust and plasma are distributed in the
same way.  Although there are reasons to suspect that this
assumption is largely valid, there are also indeed reasons to
suspect that there may be dust-plasma relative motions of a
magnitude large enough to affect our results, as we now
discuss.

In a turbulent ISM, one generically expects that--at least on
the large scales considered in this paper--dust should be
well-mixed (see, e.g., \citealt{Lazarian:2001zv}). On small
scales, however, the dust grains may not necessarily be
well coupled to the gas. From a theoretical perspective, two
major sources of coupling should be considered: collisional
coupling with the atoms in the gas, and the gyromotion of
charged grains in a magnetic field \citep{Voelk:1980,
1985prpl.conf..621D}.  The product of mean atomic velocity and
the collisional drag time in a (mostly) neutral medium is
\begin{equation}
\bar v_{\rm H} t_{\rm drag} = \frac{a\rho_{\rm g}}{m_{\rm H}n_{\rm H}}
= 5\,a_{-5}n_{\rm H}^{-1}\,{\rm pc},
\label{eq:bvt}
\end{equation}
where we have used a grain density of $\rho_{\rm g} = 2.6$
g~cm$^{-3}$, written the grain radius in units of $a_{-5} =
10^{-5}$ cm, and the hydrogen density in units of cm$^{-3}$. If
magnetic fields were neglected, we would expect the dust to
trace the gas for sound waves of (reduced) wavelength
$\lambdabar = k^{-1}$ larger than this scale. It is easily seen
that for typical ISM distances $r \sim 100$ pc, the condition of
dust-gas coupling through collisions should be violated at $\ell
= kr \gtrsim 20 a_{-5}^{-1}n_{\rm H}$, i.e.\ well within the
range of interest for the Planck dust-polarization maps.
On the other hand, the gyromotion of charged grains in
magnetic fields restricts the motion of dust grains in
directions perpendicular to the magnetic field on a length scale
of
\begin{equation}
v_{\rm A} t_{\rm L} = \frac{m_{\rm g}c}{\Phi a \sqrt{4\pi m_{\rm H}n_{\rm H}}}
= 7\times 10^{-5}\, a_{-5}^2 n_{\rm H}^{-1/2}\,{\rm pc}
\label{eq:val}
\end{equation}
for grains with a potential $\Phi\sim 10\,$V$=0.03\,$statvolt
generated by the photoelectric effect.
Therefore, for Alfv\'en waves of (reduced) wavelength
$\lambdabar = k^{-1}$ larger than this scale, we expect the dust
to trace the plasma. Factors of $\beta$ and trigonometric
factors may appear in the coupling to the slow and fast MHD
waves, but only for extreme values would we expect the Larmor
coupling to fail.

A possible exception to the above argument is that gyromotion
couples the dust to the magnetic field in the
perpendicular direction, but not in the parallel direction. To
take an extreme case, slow waves in a low-$\beta$ plasma (which
have displacements mostly along the field) with $\lambdabar$
less than Eq.~(\ref{eq:bvt}) might primarily displace the gas,
while the dust fails to participate. If the small-scale field is
itself turbulent, however, grains may undergo changes in pitch
angle and be forced to move with the gas \citep{Lazarian:2004ar}.

From an empirical perspective, the similarity of the power laws for the
dust-intensity and dust-polarization power spectra; the
difference between the E-mode and B-mode power
\citep{Adam:2014bub,Ade:2015mbc,Adam:2014gaa}; the evidence
for a similar EE/BB ratio in synchrotron radiation
\citep{Page:2006hz}; and the striking agreement of HI
21-cm and far-infrared dust maps \citep{Schlegel:1997yv} all suggest that
the dust and plasma density are not grotesquely mismatched.  If
there is indeed some random component $\delta n$, not
correlated with the magnetic-field perturbation, then that
should drive EE/BB toward unity, given the
equality of the angular averages of the $\cos2\psi$ and $\sin
2\psi$ factors that multiply $\delta n$ in
Eqs.~(\ref{eqn:Eintermediate}) and (\ref{eqn:Bintermediate}).
The observations thus suggest some correlation of the dust with
the magnetic field.  Moreover, when considering results below,
we should be looking not only for parameter combinations that
provide EE/BB~$\simeq2$, but perhaps also for those that provide
a larger ratio.

We thus proceed here under the assumption that the dust density
traces the plasma density, but note that the validity of this
assumption---and the consequences of its violation---warrant
further investigation.  Possibilities for testing the hypothesis
include the frequency dependence of the dust-polarization signal
(since dust segregation may depend on the grain size),
cross-correlation with synchrotron polarization (which is
emitted by the plasma, rather than the dust), and
cross-correlation with polarized-starlight surveys.


\section{Model of Random Displacements of the Magnetized Fluid}
\label{sec:rand}

In the previous Section we questioned whether the Planck
dust-polarization data could be explained in terms of MHD
turbulence and speculated that they might have more to do with
the large-scale turbulence-driving physics.  Here we propose a
simple phenomenological model of fluctuations of the ISM that,
as we will see, can easily produce the observed EE/BB ratio and
TE correlation.  

Instead of decomposing perturbations into slow/fast/Alfv\'en MHD
waves, we here simply suppose that the magnetized fluid
experiences a random displacement,
\begin{equation}
     {\boldsymbol{\Delta}}(\bfx) = (\Delta_1,\,\Delta_2,\,\Delta_3).
\end{equation}
The continuity equation then provides the associated density
perturbation,
\begin{equation}
    \frac{\delta n}{n_0} = -i \mathbf{k} \cdot
    \boldsymbol{\Delta} = -i k(\Delta_1 \cos\psi +
    \Delta_2\sin\psi),
\end{equation}
and from the MHD equation, $\delta\mathbf{H} = i
\mathbf{k}\times(\boldsymbol{\Delta}\times\mathbf{H}_0)$, the
associated magnetic-field perturbations are,
\begin{eqnarray}
    \frac{\delta H_x}{H_0} &=&- \tan\psi \frac{\delta H_y}{H_0}
    = -i k \Delta_2 \sin\theta \sin\psi  \\
    \frac{\delta H}{H_0} &=& i k \left( \Delta_3 \sin\theta
    \cos\theta \cos\psi  \right. \cr
    &-&\left. \Delta_1 \cos^2\theta\cos\psi - \Delta_2\sin\psi
    \right).
\end{eqnarray}
These are then inserted into
Eqs.~(\ref{eqn:Egen})--(\ref{eqn:Bgen}) to determine the E- and
B-mode polarization. To calculate the power, we assume the
displacement field has equal power in all three components,
$\langle \Delta_i \Delta_j\rangle = \delta_{ij}
F_\lambda$, where $\lambda$ now represents the anisotropy in the
displacement power.

\begin{figure}[h]
\begin{centering}
\includegraphics[width=3.25in]{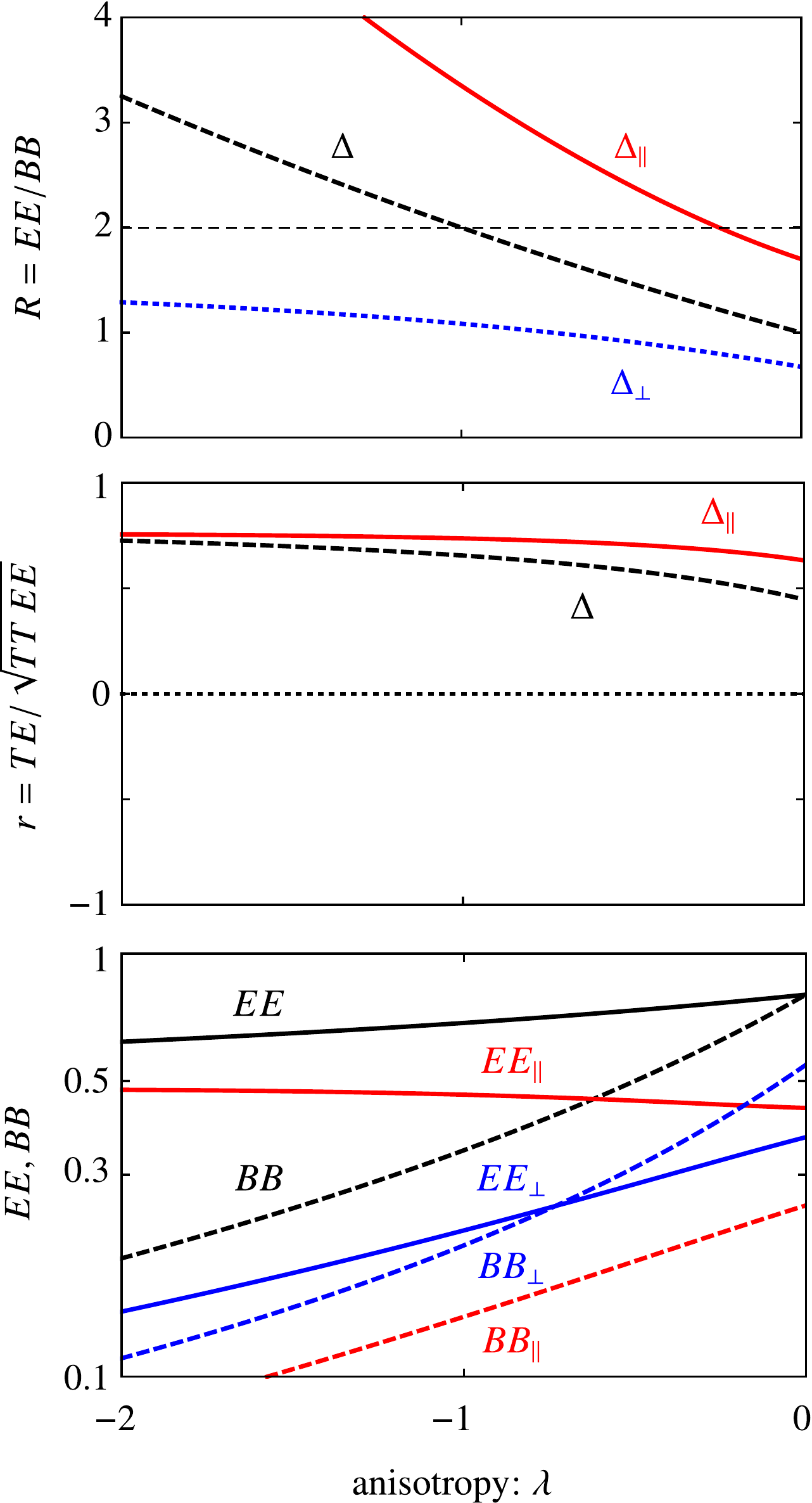}
\caption{The EE/BB ratio (top), TE cross correlation coefficient
     (middle), and E- and B-mode power normalized to the
     displacement power spectrum (bottom) are shown for  the
     model of fluid displacements (black), as well as the
     individual contributions by the longitudinal (red) and
     transverse (blue) displacements of the MHD fluid.} 
\label{fig:LTRand}
\end{centering}
\end{figure}

The results for the EE/BB ratio,  TE cross-correlation
coefficient, and individual powers are shown in
Fig.~\ref{fig:LTRand}. This model easily explains the EE/BB~$=2$
ratio and positive TE correlation with a moderately anisotropic
power index of $\lambda \simeq -1$.

To gain better insight into the physical mechanisms that could
generate a spectrum of displacements, we decompose
$\boldsymbol{\Delta}$ in the basis spanned by $\mathbf{\hat k}$,
$\mathbf{\hat a}$, and $\boldsymbol{\hat \theta}$. We define
longitudinal displacements as $\boldsymbol{\Delta}_\parallel =
(\boldsymbol{\Delta}\cdot  \mathbf{\hat k}) \mathbf{\hat k}$ and
transverse displacements $\boldsymbol{\Delta}_a$ and
$\boldsymbol{\Delta}_\theta$.  We immediately notice that the
density perturbation is entirely due to longitudinal
displacements,
\begin{equation}
     \frac{\delta n}{n_0}|_{\parallel} = -i k \Delta_{\parallel}.
\end{equation}
Hence, the observed, strong TE cross-correlation implies that
the longitudinal modes play a significant role in the structure
of the ISM on these scales. The magnetic-field fluctuations are
\begin{eqnarray}
     \frac{\delta H_x}{H_0} &=&  - \tan\psi \frac{\delta
     H_y}{H_0} = -i k \sin\psi \left(\Delta_\parallel
     \sin\alpha\sin\varpi \right. \cr 
     &&\qquad \left.  +\cos\alpha(\Delta_a  \cos\varpi -
     \Delta_\theta   \sin\varpi )\right)  \\
     \frac{\delta H}{H_0} &=& -i k \left(  \Delta_\parallel \sin^2\alpha
     - \Delta_\theta \sin\alpha\cos\alpha\right).
\end{eqnarray}
Using the above results, we can assess the relative
contributions of longitudinal- and transverse-displacement power
to the E- and B-mode power.  A similar procedure as above is
carried out to evaluate the EE/BB ratio, shown in
Fig.~\ref{fig:LTRand}. The power in transverse displacements,
indicated in the Figure as $\Delta_\perp$, consists of the sum
of $\Delta_a$ and $\Delta_\theta$ modes. In the context of this
model, the observations suggest a slightly anisotropic spectrum
of longitudinal displacements.  Although there is some
dependence of the EE/BB ratio on $\lambda$, the dependence is
relatively weak.

In addition to being fairly simple, this random-displacement
model is also fairly robust.  There is variation in EE/BB and TE
with the anisotropy parameter $\lambda$.  However, the TE
correlation is generically positive and the variation of EE/BB
with $\lambda$ fairly slow.  Clearly, this model falls far short
of a theory.  Still, it has strengths as a working model that
may help guide a more robust astrophysical explanation for the
observations.

\section{Conclusions}
\label{sec:conclusions}

We have demonstrated that the EE, BB, and TE power
spectra for polarized dust (and synchrotron) emission provide a
new, unique, and powerful probe of the state of the magnetized ISM.
We calculated the contributions to E- and B-mode power
and the TE cross-correlation from the slow, fast, and Alfv\'en
waves MHD waves and and provided results for different
ratios $\beta$ of magnetic-field to gas
pressures and different power anisotropies.  We argued that the
observations---of EE/BB/TE power and the spectral index for
fluctuations---greatly reduce the available parameter space of
MHD-turbulence models for the Planck dust-polarization data.
We then speculated that a full explanation of the observations
may involve the effects of the large-scale physics and
developed a simple phenomenological model, based on random
displacements of a magnetized fluid, that can account for the
observations.

Our work motivates a vast suite of additional
investigations.  First of all, we have used here only the
fact that the TE cross-correlation coefficient is positive.
Planck has published results for TE power, and for the TT
and EE power, but those are separate analyses that use different
cuts and assumptions about systematic effects.  It will be
valuable to measure more carefully the cross-correlation
coefficient we have calculated here.  Second, we have presented
results for EE/BB ratios and the TE cross-correlation after
averaging over all magnetic-field orientations because the
observed EE/BB~$\simeq2$ ratio seems to be quite generic across
the sky.  Still, the background-field orientation may differ
from one small patch of sky to another, and so the EE/BB ratio
and TE correlations should also vary.  If the background field
has a fixed orientation in some small patch of sky, then there
should also be a local departure from statistical isotropy within
that patch.  There is also potentially interesting
information in the $\ell$ dependence of the $C_\ell$.  Is it
really a power law?  Or does it steepen at higher $\ell$?  Are
the $\ell$ dependences of the EE, BB, TE, and TT power spectra
all the same?  Or are there subtle variations that may reflect
scale-dependent anisotropies or perhaps some other physics not
accounted for here?  We also suggest further investigation of
the frequency dependence of dust-polarization maps and 
cross-correlation with synchrotron-polarization maps and
starlight-polarization surveys to test the hypothesis that the
dust density traces the plasma density assumed here.
Finally, although we have focussed here on
Planck dust-polarization maps, similar techniques can also be
applied to dust-polarization data from specific molecular
clouds.

Fortunately, there is not only far more along these lines that
can be done with existing Planck data, but also prospects for
rich new data sets to build upon Planck.

\smallskip
We thank E.\ Vishniac for useful discussions.  MK was supported
by NSF Grant No. 0244990, NASA NNX15AB18G, the John Templeton
Foundation, and the Simons Foundation. RRC was supported by DOE
grant DE-SC0010386.  CH was supported by NASA, the U.S. Department of
Energy, the David \& Lucile Packard Foundation, and the Simons
Foundation.

\end{document}